\documentclass[amsmath,amssymb,11pt]{article}
\usepackage{jheppub2}
\pdfoutput=1

\usepackage{amsmath,amssymb,amsthm,mathrsfs,bbm,bm}
\usepackage{latexsym,amscd,amsbsy,amsfonts,dsfont}
\usepackage{multirow}
\usepackage{graphicx}
\usepackage{xcolor}
\usepackage{caption}

\newcommand{\beq}{\begin{equation}}
\newcommand{\eeq}{\end{equation}}
\newcommand{\beqa}{\begin{eqnarray}}
\newcommand{\eeqa}{\end{eqnarray}}
\newcommand{\bea}{\begin{eqnarray}}
\newcommand{\eea}{\end{eqnarray}}

\newcommand*{\affmark}[1][*]{\textsuperscript{#1}} 

\numberwithin{equation}{section}  
\allowdisplaybreaks

\title{Quantum Cross-section of Near-extremal Black Holes}

\author{Roberto Emparan\affmark[1,2]}
\emailAdd{emparan@ub.edu}

\affiliation{
\affmark[1]Institució Catalana de Recerca i Estudis Avançats (ICREA),
 Passeig Lluis Companys, 23, 08010 Barcelona, Spain\\
\affmark[2]Departament de Física Quàntica i Astrofísica and
  Institut de Ciències del Cosmos,
 Universitat de Barcelona, 08028 Barcelona, Spain\\
}

\abstract{We explore how to detect the large quantum fluctuations in the throat of a near-extremal black hole, where the dynamics are governed by the Schwarzian theory. To this end, we scatter a low-frequency wave of a massless, minimal scalar off the black hole and calculate the absorption cross-section. In the semiclassical regime, where the Schwarzian is weakly coupled, we recover the universal result that the cross-section equals the horizon area. However, in the strongly coupled regime, where quantum fluctuations dominate, we find that the absorption cross-section exceeds the semiclassical prediction.
This result may seem counterintuitive, given that the density of black hole states is suppressed in this regime. Nevertheless, two effects outweigh this suppression. First, quantum fluctuations enhance absorption transitions between individual states, with the effect becoming stronger closer to the ground state. Second, these fluctuations significantly reduce stimulated emission. We conclude that a measurement showing an enhanced absorption cross-section serves as a clear signature of the large quantum fluctuations in the geometry.
}

\begin{document}

\maketitle

\section{Introduction}
\label{sec:intro}

Is it possible to safely observe large quantum fluctuations of the spacetime geometry? This question often arises in the context of potential violations of weak Cosmic Censorship, where the semiclassical geometry breaks down near a naked singularity. However, it takes on new significance with the realization that large yet controllable quantum fluctuations of geometric quantities can occur even in regions where the curvature remains small.

It has been argued that the length of the throats of black holes sufficiently close to extremality exhibits large quantum variance \cite{Iliesiu:2020qvm}. This is because the mode controlling the connection between the mouth of the throat and the exterior geometry becomes very light when the temperature above extremality drops low enough to approach the scale 
\begin{align}
    E_b \sim \frac{1}{r_0 S_0}\,,
\end{align}
where $S_0$ denotes the (naive) semiclassical value of the extremal black hole entropy, and $r_0$ is its horizon radius.\footnote{Further studies of this effect in Reissner-Nordström black holes have been made in \cite{Iliesiu:2022onk}; for rotating Kerr and BTZ black holes, in \cite{Ghosh:2019rcj,Kapec:2023ruw,Rakic:2023vhv,Maulik:2024dwq,Kapec:2024zdj,Kolanowski:2024zrq,Maulik:2025hax}; for hyperbolic black holes, in \cite{Emparan:2023ypa}; and for holographic strange metals, in \cite{Liu:2024gxr}. We assume that the black hole is not near-BPS, and only briefly discuss the latter case in the conclusions \cite{Heydeman:2020hhw,Boruch:2022tno}.} Notably, these quantum fluctuations drastically reduce the density of black hole states when the energy above extremality, $E=M-M_0$, falls below $E_b$. With fewer available states, the Hawking emission rate is expected to be suppressed—--a result that has been explicitly confirmed in  \cite{Brown:2024ajk}.

How can an external observer probe this large quantum object?\footnote{To avoid any misunderstanding, we stress that we are posing a question of principle explored through idealized thought experiments, undeterred by constraints from Standard Model matter or foreseeable technology.} One possible approach involves using Hawking radiation as the primary means. In this scenario, one would first collapse charged matter to form a black hole initially outside the quantum near-extremal regime. The black hole would then be allowed to evaporate through Hawking emission until it approaches extremality. At this stage, the radiation begins to deviate noticeably from the predictions of the  semiclassical picture, revealing the effects of the underlying quantum geometry.

The main issue with this approach, however, is its inordinate slowness \cite{Brown:2024ajk}. The entire process—both the approach to extremality and the subsequent near-extremal emission—is driven by quantum dynamics, with timescales that diverge as $\hbar \to 0$, which makes them exceedingly long for macroscopic black holes. The slowness is further exacerbated by the suppression of the emission rates in the regime of interest.

A more efficient strategy is to investigate the quantum black hole as classical experimentalists: using a classical mechanism for forming the near-extremal black hole, and observing its properties through classical means. Recent work has demonstrated that the collapse of charged matter can be fine-tuned to produce a black hole that, in classical terms, would be exactly extremal \cite{Kehle:2024vyt}. In our context, where a classical extremal black hole does not exist, this result suggests (and it would be interesting to examine this further) that the collapse can be similarly fine-tuned to create a black hole very close to the quantum near-extremal regime, with an energy within the range of $E_b$.\footnote{Fine-tuning is, of course, a standard experimental practice. If this case is analogous to Choptuik’s critical collapse, the fine-tuning required will scale as a power law of the initial data, rather than exponentially, making it relatively moderate.}
Since the collapse is governed by classical gravity, it will happen in a timescale very close to (within $O(\hbar)$ of) the time for collapse to extremality. In other words, the near-extremal black hole forms rapidly, within a timescale on the order of its radius. This makes it a significantly faster and more efficient method for accessing the quantum regime of near-extremal black holes.

To observe this black hole, we propose conducting a scattering experiment: sending a wave toward the black hole and measuring its absorption cross-section. This process is relatively quick and, as we will show, can yield results that differ significantly from the semiclassical prediction.

But what kind of scattering experiment should we perform? Using null geodesics of photons is not a good idea. First, it risks destroying the long throat and the quantum fluctuations we aim to study. Second, and more critically, these geodesics only probe the local geometry of the throat, which does not exhibit strong fluctuations. Simply shining light rays on the quantum black hole would produce a classical image, failing to reveal the quantum effects of interest.

Instead, we need to gently probe the mouth of the throat. To achieve this, we will scatter a minimally-coupled massless scalar field off the black hole. In the very low-frequency limit, the scattering is dominated by the s-wave component, which works to our advantage: s-waves probe the homogeneous fluctuations in the region near the horizon.

Using low frequencies is also convenient because it helps ensure that the process remains within the quantum regime without destroying the throat. However, the frequency must not be too low; otherwise, instead of effectively probing the quasi-continuous distribution of states, it would become sensitive to the fundamental discreteness of the black hole spectrum---a fascinating prospect but beyond current computational control. We will provide concrete constraints on the parameter ranges under which these conditions are satisfied.

The low-frequency scattering of massless minimal scalars off a classical black hole is a well-studied topic, initiated in \cite{Starobinsky:1973aij,Ford:1975tp,Unruh:1976fm} and significantly developed in the years leading up to the AdS/CFT correspondence (e.g, \cite{Das:1996wn,Gubser:1996xe,Maldacena:1996ix,Das:1996we,Gubser:1996zp,Callan:1996tv,Maldacena:1997ih}). In the limit $\omega\to 0$, the absorption cross-section for a spherically symmetric black hole is equal to the horizon area \cite{Das:1996we}
\begin{equation}\label{univsigma}
    \sigma_{abs}=A_H \qquad \textrm{(semiclassical)}\,,
\end{equation}
as expected from s-wave dominance in this regime. This universal result serves as a benchmark for comparing our findings on the quantum absorption cross-section.
Furthermore, it suggests that $\sigma_{abs}$ provides a measure of the density of absorbing states of the black hole, $\rho(E)$, in the semiclassical regime. Using the Bekenstein-Hawking entropy formula, we can rewrite \eqref{univsigma} as
\begin{equation}
    \sigma_{abs}=4G \log \rho(E) \qquad \textrm{(semiclassical)}\,.
\end{equation}

When the scalar field probes a near-extremal black hole, it becomes sensitive to the quantum fluctuations of the throat and the quantum nature of the black hole states.
Given that quantum fluctuations reduce the density of states, with $\rho(E)\ll e^{A_H/4G}$, one might expect that the quantum absorption cross-section would be much smaller than the classical horizon area. However, as we will demonstrate, this expectation is not correct. We find that the low-frequency scattering of a scalar wave off a quantum throat results in an enhancement of the absorption cross-section,
\begin{equation}
    \sigma_{abs}>A_H>4G \log \rho(E)  \qquad \textrm{(quantum)}\,.
\end{equation}
Indeed, $\sigma_{abs}$ increases as the black hole approaches extremality, even though the black hole has fewer states. This may seem a counterintuitive result, but it arises because quantum fluctuations introduce two effects that more than compensate for the suppression of states. The primary effect is the enhancement of absorption transitions between individual states. The second, less significant but still relevant, is the suppression of stimulated emission.

Thus, the answer to the question we posed at the beginning is affirmative: classical experiments can probe the quantum regime of near-extremal black holes. Since the value of $A_H$ can be inferred by, e.g., measuring the black hole charge, an enhanced absorption cross-section will reveal the large quantum fluctuations of the throat.

\bigskip 

\noindent\emph{Note:} While we were writing up our results, we became aware of upcoming work by Anna Biggs, where, among other things, the quantum absorption cross-section is computed \cite{Biggs:2025nzs}. That work should be given equal credit for independently performing the study in our article. 

\section{Quantum absorption}

For simplicity we will present our analysis for a four-dimensional Reissner-Nordström black hole, but we will later see that the results easily generalize to other static, spherically-symmetric black holes in any dimension. Close to extremality, the semiclassical black hole mass is
\begin{equation}\label{MofT}
    M(T)=M_0 +2\pi^2 \frac{T^2}{E_b}+O(T)^3\,,
\end{equation}
with\footnote{The subindex $b$ is for `breakdown' of the $SL(2,R)$ symmetry of AdS$_2$.}
\begin{equation}\label{Eb2}
    E_b=\frac{\pi}{r_0 S_0}.
\end{equation}
As argued in \cite{Preskill:1991tb}, for $T\lesssim E_b$ the semiclassical thermodynamics breaks down. Following \cite{Iliesiu:2020qvm}, we now understand that, as in conventional laboratory systems, so too in black holes, quantum effects become paramount at low temperatures.
This follows from a careful study of the relevant sector of black hole dynamics in this regime, which reduces the system to a simple one-dimensional Schwarzian theory living at the boundary of the AdS$_2$ throat. We refer the reader to \cite{Moitra:2019bub,Iliesiu:2020qvm,Iliesiu:2022onk,Mertens:2022irh} for details of this reduction.

\subsection{Quantum black hole interacting with classical scalar field}

We propose to send a wave of a massless, minimally coupled scalar field into a near-extremal black hole, characterized by the parameter $E_b$ and its initial energy above extremality,
\begin{equation}\label{Einext}
    E_i=M-M_0\,.
\end{equation}
The wave is modeled as a coherent state of the field with a large occupation number, $\langle N_\omega\rangle\gg 1$ and a frequency 
\begin{equation}\label{lowbound}
    \omega >E_b\, e^{-S_0}\,.
\end{equation}
This condition ensures that the wave interacts with a quasi-continuous spectrum of states. Our main interest will be on the regime where $\omega$ and $E_i$ are of the order of or smaller than $E_b$. If $\omega > E_i$, the black hole cannot emit radiation of this frequency, but it remains capable of absorbing it. We restrict our analysis to the s-wave sector, though higher partial waves could also be studied.

The phenomenon we examine closely resembles early studies of a quantum atom interacting with a classical electromagnetic field. The coupling of a massless, minimal scalar to the quantum Schwarzian mode of a black hole was recently derived in \cite{Brown:2024ajk}, and we will largely build on their analysis, keeping our presentation concise.
The key distinction lies in the initial state of the scalar field. While \cite{Brown:2024ajk} focused on Hawking-like spontaneous emission, where the scalar field is initially in its ground state, we instead consider it in a semiclassical configuration. This setup allows for both absorption by the black hole and stimulated emission.

Since the field is neutral and restricted to the s-wave sector, the charge and angular momentum of the black hole remain fixed, allowing us to treat it within the microcanonical ensemble. While it would be possible to study a thermal radiation state for the field and analyze the black hole in the canonical ensemble, we do not pursue that here \footnote{Absorption and emission in the canonical ensemble were studied in \cite{Bai:2023hpd}.}.

We consider a scalar wave with ingoing and outgoing components at infinity,
\begin{equation}
    \Psi(t,r)\sim c_{in} e^{-i\omega (t-r)}+c_{out} e^{-i\omega (t+r)}\,,
\end{equation}
which propagates classically to the mouth of the black hole throat. There, the wave acts as a source providing a boundary condition for the effective 2D field near the boundary of the AdS$_2$ region,
\begin{equation}
    \Psi(t,r)=\psi(t) r^{\Delta-1}+O(r^{-\Delta})\,,
\end{equation}
 where $\Delta$ is the conformal weight of the dual operator. For the s-wave of a massless, minimal scalar field, $\Delta=1$.

The relevant part of the action for the coupled system of the black hole and the scalar wave simplifies to
\begin{equation}
    I=I_{Schwarzian}+\int dt\, \psi(t) \mathcal{O}(t)\,,
\end{equation}
where the Schwarzian term describes the quantum states of the black hole, and
$\mathcal{O}(t)$ is the operator that measures the response to $\psi(t)$, acting as its conjugate.
This is a familiar expression in AdS/CFT holography, where $\mathcal{O}$ corresponds to the operator dual to the massless scalar field in AdS$_2$. 

The scalar field wave we send to the black hole,
\begin{equation}
    \psi(t)=\psi_0\, e^{-i\omega t}
\end{equation}
acts as an oscillating source at the mouth of the throat, exciting the black hole. The response of the black hole is encoded in the matrix elements of the operator 
$\mathcal{O}$, which appears in the interaction Hamiltonian
\begin{equation}
    H_{int}=e^{-i\omega t}\,\psi_0\, \mathcal{O}(t)\,.
\end{equation}
This interaction drives transitions between black hole quantum states,
\begin{align}
    |E_i\rangle\to |E_i +\omega\rangle &\qquad \textrm{absorption}\,,\\
    |E_i\rangle\to |E_i -\omega\rangle &\qquad \textrm{emission~(stimulated)}\,.
\end{align}

\subsection{From transition rates to absorption cross-section}

To compute the transition rates between the initial and final states of the black hole–radiation system, we use Fermi's Golden Rule,
\begin{equation}
    \mathcal{T}_{i,f}=2\pi|\langle E_f,N_f|\mathcal{O}\psi_0|E_i,N_i\rangle|^2 \rho(E_f)\,,
\end{equation}
where $\rho(E_f)$ is the density of final states. For simplicity, we now treat the radiation states as eigenstates of the occupation number rather than coherent states; this distinction becomes negligible in the regime of interest.
Assuming the occupation number of the wave is large, such that $|N_i-N_f|\ll N_i=N_\omega$, we can neglect spontaneous emission (Hawking radiation) since $N_\omega+1\simeq N_\omega$. 

The proper normalization of the amplitude $\psi_0$ is determined through a matching calculation. This involves matching the field at the mouth of the throat with the wave in the asymptotically flat region, where it is normalized to ensure that the in- and out-amplitudes (quantized as annihilation operators) satisfy standard commutation relations.\footnote{This is related to, but not identical to, the classical calculation of the absorption cross-section using matched asymptotics.} This normalization was performed in \cite{Brown:2024ajk}, yielding\footnote{We note a factor of two discrepancy with their value of $\mathcal{N}^2$.}
\begin{equation}
    |\psi_0|^2=\langle N_\omega\rangle \frac{r_0^2\omega}{\pi^2}\,.
\end{equation}
Under the assumption of a large occupation number for the wave, $\langle N_\omega\rangle$ factors out of both the emission and absorption transition rates
\begin{equation}
    \mathcal{T}_{i,f}\propto \langle N_\omega\rangle\,.
\end{equation}
This highlights one of the advantages of the classical scattering experiment as a much more efficient probe of the black hole than Hawking emission. While the transition rates between individual states are the same in both cases, the classical wave stimulates them with a large number $\langle N_\omega\rangle\gg 1$ of external quanta.

Taking into account a delta-function $\delta(E_f-E_i\pm \omega)$ , which enforces energy conservation, and integrating over final energies, we can express the absorption and emission rates per unit frequency as\footnote{These rates were computed in the 2D theory of the throat, without coupling to the asymptotic region, in \cite{Mertens:2019bvy,Blommaert:2020yeo}.}
\begin{align}\label{Gammas}
    \Gamma_{abs}(\omega)&=\langle N_\omega\rangle \frac{2 r_0^2\omega}{\pi}|\langle E_i+\omega|\mathcal{O}|E_i\rangle|^2 \rho(E_i+\omega)\\
    \Gamma_{emit}(\omega)&=\langle N_\omega\rangle \frac{2 r_0^2\omega}{\pi}|\langle E_i-\omega|\mathcal{O}|E_i\rangle|^2 \rho(E_i-\omega)\\
    &=-\Gamma_{abs}(-\omega)\,.
\end{align}
The final equality reflects time-reversal invariance.

When calculating the total absorption rate per mode, it is essential to account for the fact that stimulated emission must be subtracted, that is,
\begin{equation}
    \frac{d\langle N_\omega\rangle}{dt d\omega}=-\left(\Gamma_{abs}(\omega)-\Gamma_{emit}(\omega)\right)\,.
\end{equation}
The total absorption rate is then expressed in terms of the absorption probability,  $P_{abs}(\omega)$, as \cite{Page:1976df}
\begin{equation}
    \frac{d\langle N\rangle}{dt}=-\int \frac{d\omega}{2\pi} P_{abs}(\omega)\langle N_\omega\rangle\,,
\end{equation}
where
\begin{align}
    P_{abs}(\omega)=\frac{2\pi}{\langle N_\omega\rangle}\left(\Gamma_{abs}(\omega)-\Gamma_{emit}(\omega)\right)\,.
\end{align}
This quantity is often referred to as the `greybody factor'.

To convert the absorption probability into the absorption cross-section, we employ the optical theorem. For a massless, minimally coupled scalar in four dimensions, the relationship is given by \cite{Das:1996we,Gubser:1996zp}
\begin{align}
    \sigma_{abs}(\omega)&=\frac{\pi}{\omega^2}P_{abs}(\omega)\\
    &=\frac{2\pi^2}{\omega^2\langle N_\omega\rangle}(\Gamma_{abs}(\omega)-\Gamma_{emit}(\omega))\,.
\end{align}
Substituting \eqref{Gammas} into this, we arrive at
\begin{equation}\label{sigma1}
    \sigma_{abs}=\frac{4\pi r_0^2}{\omega}\left( |\langle E_i+\omega|\mathcal{O}|E_i\rangle|^2 \rho(E_i+\omega) - |\langle E_i-\omega|\mathcal{O}|E_i\rangle|^2 \rho(E_i-\omega)\right)\,.
\end{equation}

\subsection{Quantum absorption cross-section}

The density of black hole states and the matrix elements of conformal primaries required for \eqref{sigma1} have been explicitly computed in comprehensive studies of the Schwarzian theory and its coupling to conformal matter \cite{Stanford:2017thb, Mertens:2017mtv,Kitaev:2018wpr,Yang:2018gdb,Iliesiu:2019xuh, Mertens:2022irh}. The density of states is given by
\begin{equation}
    \rho(E)=\frac{e^{S_0}}{2\pi^2 E_b}\sinh\left(2\pi\sqrt{2E/E_b}\right)\Theta(E)\,.
\end{equation}
The matrix elements of the marginal operator $\mathcal{O}$ with $\Delta=1$ are extracted from the Fourier transform of the corresponding two-point function, yielding
\begin{equation}
    |\langle E_f|\mathcal{O}|E_i\rangle|^2=2\pi^2 E_b\, e^{-{S_0}}\frac{E_f-E_i}{\cosh\left(2\pi\sqrt{2E_f/E_b}\right)-\cosh\left(2\pi\sqrt{2E_i/E_b}\right)}\,.
\end{equation}

Substituting these into \eqref{sigma1}, we arrive at our main result,
\begin{align}\label{qcross}
    \sigma_{abs}=A_H\left(\frac{\sinh\left(2\pi\sqrt{\frac{2(E_i+\omega)}{E_b}}\right)}{\cosh\left(2\pi\sqrt{\frac{2(E_i+\omega)}{E_b}}\right)-\cosh\left(2\pi\sqrt{\frac{2E_i}{E_b}}\right)}-
    \frac{\sinh\left(2\pi\sqrt{\frac{2(E_i-\omega)}{E_b}}\right)\Theta(E_i-\omega)}{\cosh\left(2\pi\sqrt{\frac{2E_i}{E_b}}\right)-\cosh\left(2\pi\sqrt{\frac{2(E_i-\omega)}{E_b}}\right)}
    \right),
\end{align}
where the semiclassical black hole area has been identified as
\begin{equation}
    A_H=4\pi r_0^2\,.
\end{equation}
In this expression, the first term corresponds to absorption transitions that increase the energy of the black hole, while the second term accounts for emission transitions that decrease the energy and reduce the total absorption. This second term vanishes when $\omega\geq E_i$, as emission is impossible in this case.

The semiclassical approximation to the black hole holds when $E_i\gg \omega, E_b$, and in this limit we expect the transition rates to approximate their thermal values. Expanding for small $E_b/E_i$ and $\omega/E_i$ we obtain
\begin{align}
     |\langle E_i+\omega|\mathcal{O}|E_i\rangle|^2 \rho(E_i+\omega)&\to \frac{\omega\, e^{\omega/T}}{e^{\omega/T}-1}\,,\\
    |\langle E_i-\omega|\mathcal{O}|E_i\rangle|^2 \rho(E_i-\omega)&\to \frac{\omega}{e^{\omega/T}-1}\,,
\end{align}
where
\begin{equation}
    T=\frac{\sqrt{2E_i E_b}}{2\pi}
\end{equation}
is the Hawking temperature of the black hole, as can be seen from \eqref{MofT} and \eqref{Einext}. In this regime, the factor in brackets in \eqref{qcross} simplifies to 1, and we recover the semiclassical absorption cross-section
\begin{equation}
    \sigma_{abs}\to A_H\,.
\end{equation}
\begin{figure}[t]
\begin{center}
    \includegraphics[width=0.9\textwidth]{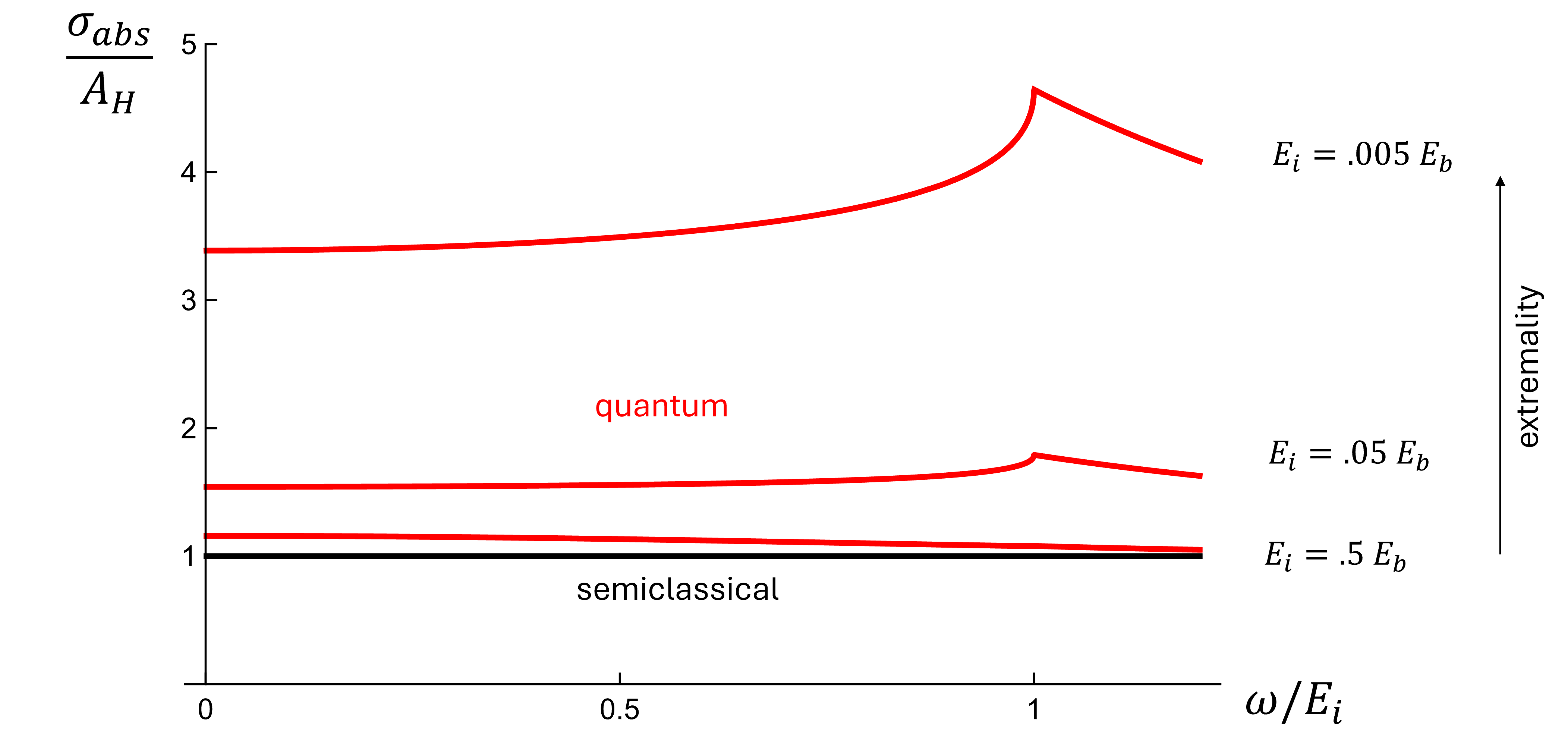}
\end{center}
\caption{\small Ratio $\sigma_{abs}/A_H$ as a function of the frequency $\omega$ for different values of the energy $E_i$ above extremality, cf.~\eqref{qcross}. The black line is the semiclassical value $\sigma_{abs}/A_H=1$, which is approached when $E_i\gg E_b$.  For small $E_i/E_b$, the zero-frequency limit is $\approx \frac1{\pi}\sqrt{\frac{E_b}{2E_i}}$, cf.~\eqref{largesigma}. The absorption grows with $\omega$ because stimulated emission with decay into lower energy states is increasingly suppressed, until emission is no longer possible for $\omega\geq E_i$, which gives the kink in the curves.} 
  \label{fig:sigmaA}
\end{figure}

We are interested in the deviations from the semiclassical limit when the energy $E_i$ of the black hole is comparable to $E_b$. Numerical evaluation easily shows that the term in brackets in \eqref{qcross} is always greater than one, as illustrated in Fig.~\ref{fig:sigmaA}. Therefore, the absorption cross-section is enhanced
\begin{equation}
    \sigma_{abs}> A_H\,,
\end{equation}
confirming our earlier statement in the introduction.

Simple analytical expressions can be found in the limit of very low frequency with $E_i/E_b$ fixed. The result \eqref{qcross} simplifies to
\begin{equation}
    \sigma_{abs}(\omega\to 0)\to A_H \left(\coth\left(2\pi\sqrt{\frac{2E_i}{E_b}}\right)+\frac1{2\pi}\sqrt{\frac{E_b}{2E_i}}\right) \,,
\end{equation}
which is manifestly greater than $A_H$ since $\coth >1$. When $E_b\ll E_i$, we find small corrections to the semiclassical value
\begin{equation}
    \sigma_{abs}(\omega\to 0)\simeq A_H\left(1+\frac1{2\pi}\sqrt{\frac{E_b}{2E_i}}+ 2 \exp\left(-4\pi\sqrt{\frac{2E_i}{E_b}}\right)+\dots\right)\,.
\end{equation}

On the other hand, close to the ground state, $E_i\ll E_b$, the absorption cross-section increases dramatically
\begin{equation}\label{largesigma}
    \sigma_{abs}(\omega\to 0)\simeq A_H\frac1{\pi}\sqrt{\frac{E_b}{2E_i}}\gg A_H\,.
\end{equation}
This may seem surprising. In the introduction, we noted that the semiclassical absorption cross-section might be viewed as a proxy for the number of black hole states. However, deep in the quantum regime we find that
\begin{equation}
    \sigma_{abs}\gg 4G S_0 \gg 4G \log \rho(E_i)\,,
\end{equation}
so the density of states significantly underestimates the actual absorption cross-section. In other words, although the black hole may have only a few states available for radiation absorption, these states are highly efficient at it. This is actually expected: the enhancement of the transition rates mediated by  $\mathcal{O}$ is directly related to the late-time behavior of its two-point function, which is significantly larger than the exponential quasinormal decay predicted by semiclassical black hole physics, and instead falls off as a power law $\sim t^{-3/2}$.

\begin{figure}[t]
\begin{center}
    \includegraphics[width=\textwidth]{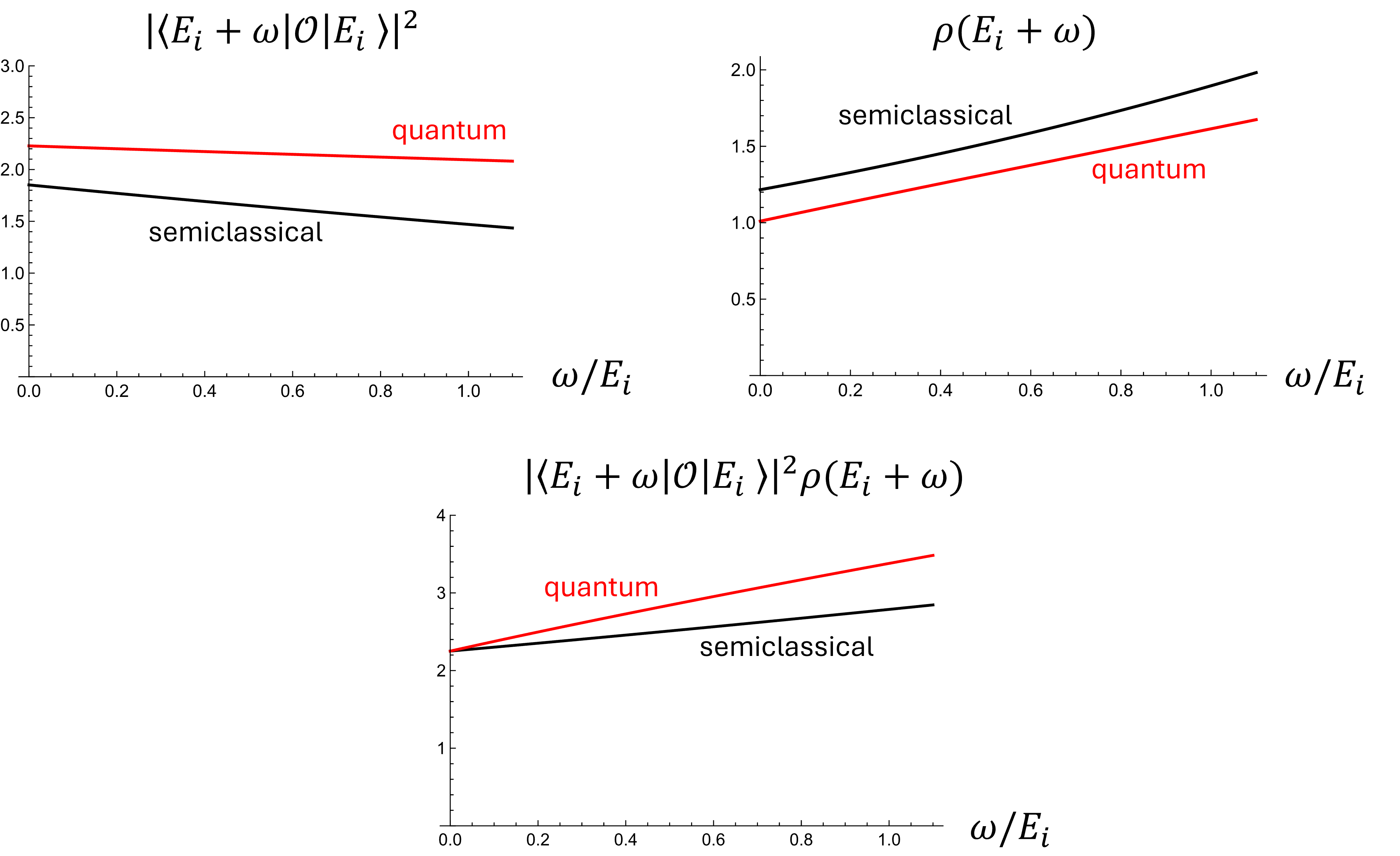}
\end{center}
\caption{\small Quantum vs.~semiclassical absorption transitions as a function of $\omega$ for $E_i=0.1 E_b$. While the density of final states, $\rho(E_i+\omega)$, is suppressed, the absorption transitions into individual states, $|\langle E_i+\omega|\mathcal{O}|E_i\rangle|^2$, are strongly enhanced by quantum fluctuations. As a result, the total absorption rate is ultimately increased. (Units in the vertical axis are arbitrary).
} 
  \label{fig:absorb}
\end{figure}
\begin{figure}[t]
\begin{center}
    \includegraphics[width=\textwidth]{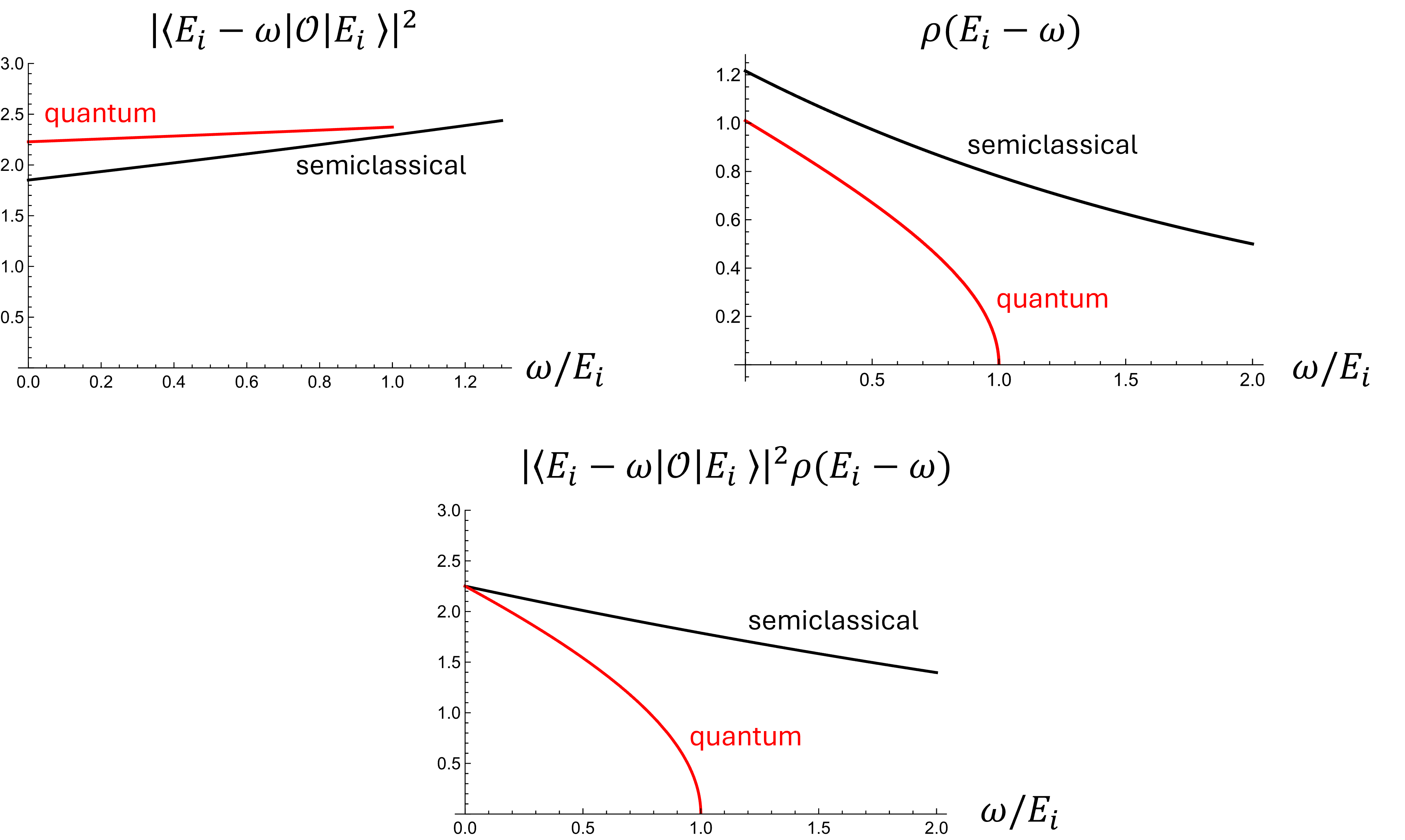}
\end{center}
\caption{\small Quantum vs.~semiclassical emission transitions as a function of $\omega$ for $E_i=0.1 E_b$. Although the allowed emission transitions for $\omega<E_i$, characterized by $|\langle E_i-\omega|\mathcal{O}|E_i\rangle|^2$, are enhanced by quantum fluctuations,  the suppression of the density of final quantum states, $\rho(E_i-\omega)$, is much stronger. Consequently, the (stimulated) emission rate is suppressed. (Units in the vertical axis are arbitrary, but the same as in Fig.~\ref{fig:absorb}.).
} 
  \label{fig:emit}
\end{figure}
This observation is illustrated in Figures~\ref{fig:absorb} and \ref{fig:emit}. In Fig.~\ref{fig:absorb} we see that although the density of absorbing states diminishes at lower energies ($E_i < E_b$), the enhancement in individual absorption transitions compensates for this reduction, leading to an overall increase in absorption. On the other hand, in Fig.~\ref{fig:emit} we observe that the competing effect of stimulated emission is suppressed by the density of final states. As the black hole transitions to a lower energy state, the smaller number of available final states significantly reduces the emission rates.

\subsection{No throat disruption}

Since the absorption probability is enhanced, one may worry that the absorption by the black hole could be large enough to disrupt the quantum throat. For this to happen, the total energy absorbed by the black hole must approach $E_b$. 

The absorption rate of energy is given by
\begin{align}
    \frac{dE}{dt d\omega}&=\omega \left(\Gamma_{abs}(\omega)-\Gamma_{emit}(\omega)\right)\\
    &=\frac{\omega^3}{2\pi}\langle N_\omega\rangle \sigma_{abs}\,.
\end{align}
The longer the scattering experiment takes, the larger the absorption. At a minimum, it must last a time $\Delta t\gtrsim 1/\omega$. The energy absorbed by the black hole in this time is
\begin{equation}
    \Delta E\gtrsim \omega^3 \langle N_\omega\rangle \sigma_{abs}\,.
\end{equation}

It is easy to verify that, close to extremality, the maximum of $\sigma_{abs}$ at $\omega=E_i$ is only $\sqrt{2}$ larger than the zero-frequency value \eqref{largesigma}. Therefore, a good parametric estimate of the absorption cross-section is 
\begin{equation}
    \sigma_{abs}\sim A_H \sqrt{\frac{E_b}{E_i}} \sim r_0^2 \sqrt{\frac{E_b}{E_i}}\,.
\end{equation}
If we require that $\Delta E< E_b$ to not disrupt the throat, then the occupation number of the scalar wave must not exceed the limit
\begin{equation}
     \langle N_\omega\rangle < \frac1{(\omega r_0)^2}\frac{\sqrt{E_b E_i}}{\omega}\,.
\end{equation}

Our aim is to probe black holes in the range of $E_i\lesssim E_b \sim 1/(r_0 S_0)$. Then we need
\begin{equation}
     \langle N_\omega\rangle < \frac1{(\omega r_0)^3 S_0}\,.
\end{equation}
The low-frequency condition underlying the classical analysis is that $\omega r_0\ll 1$. We want waves of even lower frequencies, $\omega \lesssim E_b \sim 1/(r_0 S_0)$. In this case, the black hole will remain in the quantum regime if 
\begin{equation}
    \langle N_\omega\rangle < S_0^2\,.
\end{equation}
Since $S_0\gg 1$ this easily allows for large occupation numbers. Therefore, it is well possible to probe the quantum regime of black holes with $E_i<E_b$ using semiclassical waves of frequencies comparable or smaller than $E_b$.

\subsection{Generality and universality of the quantum cross-section}

The derivation of the quantum cross-section \eqref{qcross} relies on only a few assumptions. The black hole must be static and have an AdS$_2$ throat near the horizon. The boundary mode must be a Schwarzian theory, which occurs when the gravitational sector of the theory is Einstein-Hilbert. There should not be other light modes contributing, which is generally the case except when supersymmetry is present.\footnote{As we mentioned, we freeze the light rotational and electromagnetic modes by keeping the black hole spin and charge fixed.}

Under these conditions, the quantum near-extremal system will be characterized by $A_H$ and $E_b$, with a quasi-continuous spectrum of black holes with energy $E$ above the ground state. In the semiclassical limit, the absorption cross-section must equal $A_H$. The only parameter of the Schwarzian theory is $E_b$, defined as in \eqref{MofT}. Therefore the quantum cross-section can be obtained by replacing the corresponding values of $A_H$ and $E_b$ in \eqref{qcross}. For instance, for the Reissner-Nordström solution in $D$ spacetime dimensions, we must use
\begin{equation}
    E_b=\frac{(D-3)^2}{D-2}\frac{2\pi}{r_0 S_0}\,, \qquad A_H=\Omega_{D-2} r_0^{D-2}\,.
\end{equation}

\section{Final remarks}

\paragraph{Absorbing other fields.} We have focused on the absorption of s-wave, minimal massless scalars, as they dominate at low frequencies and are the most effective in probing the Schwarzian fluctuations. However, one could also study the absorption of higher partial waves, as well as photons and gravitons, giving rise to qualitatively new effects. For instance, the absorption of radiation with spin $j\geq 1$ raises the black hole energy above $E_b$ and changes it into a rotating black hole (the spontaneous emission of these fields has been studied in \cite{Brown:2024ajk}). However, already at the classical level the absorption cross-section of these fields vanishes with a power of the frequency \cite{Mathur:1997et,Gubser:1997qr}, and it is not immediately clear whether quantum fluctuations will enhance or suppress absorption. 

One may question our thought experiment, given the apparent absence of exactly massless scalars in our universe. However, it is possible that a yet-undetected ultra-light scalar field exists. If its mass is greater than $E_b$, then its absorption would destroy the long throat, but if it is lighter (and lighter than $E_i$), it could be used for our classical scattering experiment.

\paragraph{Near-BPS black holes.} When the black hole is not only near-extremal but also near-supersymmetric, additional contributions to the light-mode fluctuations significantly alter the spectrum, opening a gap of size $\sim E_b$ between the degenerate supersymmetric ground state and the quasi-continuous spectrum of states at finite temperature \cite{Heydeman:2020hhw,Boruch:2022tno}.  To the extent that the density of states in the gap exactly vanishes, no black holes exist within this energy range. In this case, absorption by the BPS ground state will not occur until the gap is bridged. Similarly, a black hole initially above the gap will be able to emit quanta only to final states outside the gap. Studying these processes is an interesting challenge but requires additional work.

\paragraph{BTZ near horizon.} The greybody factors for near-extremal black strings with a BTZ geometry near the horizon encode information about the left and right movers of the dual CFT$_2$ \cite{Maldacena:1996ix, Gubser:1996zp}. However, in the near-extremal limit described by the Schwarzian theory this structure simplifies to the same form we have studied here. Note that the Schwarzian sector of the CFT$_2$ has been identified in \cite{Ghosh:2019rcj}.

\paragraph{Kerr black hole.} The study of greybody factors for the Kerr black hole is more involved than for static black holes. However, the interaction of the Schwarzian theory with massless fields has recently been analyzed in \cite{Maulik:2025hax}, suggesting that extending our study to the Kerr black hole is possible.

\paragraph{Quantum $\eta/s$?} The classical universal result $\sigma_{abs}/A_H \to 1$ at low frequencies famously implies the universal value $\eta/s = 1/4\pi$ in the hydrodynamics of holographic plasmas \cite{Policastro:2001yc, Kovtun:2004de}. This connection arises because shear modes on the worldvolume of a black brane behave like minimal scalars, with their absorption leading to viscous dissipation on the brane, both processes determined by the two-point function of the stress-energy tensor. An intriguing question, then, is whether our results imply that low-temperature, near-extremal holographic plasmas satisfy
\begin{equation}
\left(\frac{\eta}{s}\right)_{quantum}>\frac1{4\pi}\,\textrm{?}
\end{equation}

The issue, however, is delicate. At first glance, the hydrodynamic regime requires wavelengths longer than the thermal length, $1/T$, implying that the spatial extent of the brane worldvolume must also be larger than $1/T$. This makes the horizon very long, which, in turn, causes the scale $E_b$ to become very small. The reason is that the Schwarzian mode uniformly shifts the entire brane horizon, so when the horizon is very large, its quantum fluctuations are small, meaning the temperature range where quantum effects are relevant is pushed to extremely low values.

For example, consider a near-extremal Reissner-Nordström-AdS$_4$ black brane \cite{Chamblin:1999tk}. In this case,
\begin{equation}
    E_b\sim \frac{\ell_p^2}{L^2 r_0}\,,
\end{equation}
where $\ell_p$, $L$, and $r_0$ denote the Planck length, the AdS radius, and the horizon radius, respectively. To stay within the naive hydrodynamic regime, $T > 1/L$, while also maintaining $T < E_b$ to ensure significant quantum fluctuations, one would require $L r_0 < \ell_p^2$. But this would place the geometry in the sub-Planckian regime, where we no longer have a valid description.

However, it has been suggested that the gradient expansion of holographic hydrodynamics may not be limited by $1/T$, but rather by a length scale that at low temperatures would be much shorter, $\sim L^2 / r_0$ \cite{Davison:2013bxa}. If this is the case, hydrodynamics might apply for near-extremal black branes with $r_0 > L$, allowing access to the strongly coupled Schwarzian dynamics at temperatures that, while still low, remain above the discreteness scale $E_b\, e^{-S_0}$. Determining whether this is a viable possibility requires further investigation.

\section*{Acknowledgments}

I am grateful to Luca Iliesiu for his comments on the draft, and to Anna Biggs for conversations where we agreed on how to release our results. This work has been supported by MICINN grant PID2022-136224NB-C22, AGAUR grant 2021 SGR 00872, and State Research Agency of MICINN through the ``Unit of Excellence María de Maeztu 2020-2023'' award to the Institute of Cosmos Sciences (CEX2019-000918-M).

\bibliography{QCSbib}

\end{document}